\documentclass[12pt,preprint]{aastex}
\usepackage{psfig,emulateapj5}

\shorttitle{Eight New Extremely Metal-Poor Galaxies}
\shortauthors{Kniazev et al.}

\begin{document}

\slugcomment{Accepted to ApJ Letters}

\title{Discovery of Eight New Extremely Metal--Poor Galaxies in the Sloan
Digital Sky Survey}

\author{Alexei Y. Kniazev\altaffilmark{1,3}, Eva K. Grebel\altaffilmark{1},
Lei Hao\altaffilmark{2}, Michael A. Strauss\altaffilmark{2},
Jonathan Brinkmann\altaffilmark{4}, Masataka Fukugita\altaffilmark{5}}

\altaffiltext{1}{Max-Planck-Institut f\"{u}r Astronomie, K\"{o}nigstuhl 17,
D-69117 Heidelberg, Germany; kniazev@mpia.de}
\altaffiltext{2}{
Princeton University Observatory, Peyton Hall, Princeton, NJ 08544-1001}
\altaffiltext{3}{Special Astrophysical Observatory, Nizhnij Arkhyz,
      Karachai-Circessia, 369167, Russia}
\altaffiltext{4}{Apache Point Observatory, P.O. Box 59, Sunspot, NM 88349}
\altaffiltext{5}{Institute for Cosmic Ray Research, University of Tokyo,
      5-1-5 Kashiwa, Kashiwa City, Chiba 277-8582, Japan}

\email{kniazev@mpia.de, grebel@mpia.de,\\
haol@astro.princeton.edu, strauss@astro.princeton.edu,\\
jb@apo.nmsu.edu, fukugita@icrr.u-tokyo.ac.jp}

\begin{abstract}
We report the discovery of eight new extremely metal-poor  galaxies
(XMPGs; 12+log(O/H)~$<$~7.65)
and the recovery of four previously known or suspected XMPGs
(I~Zw~18, HS~0822$+$3542, HS~0837$+$4717 and A1116$+$517)
using Sloan Digital Sky Survey (SDSS) spectroscopy.
These new objects were identified after an analysis of 250,000 galaxy
spectra within an area of $\sim$3000 deg$^2$ on the sky.
Our oxygen abundance determinations have an
accuracy of $\le$ 0.1 dex and are based on the temperature-sensitive
[O\,{\sc iii}] $\lambda$4363 \AA\ line and on the direct calculation
of the electron temperature.  
We briefly discuss a new method of oxygen abundance determinations using
the [O\,{\sc ii}] $\lambda$7319,7330 \AA\ lines, which is particularly 
useful for SDSS emission-line spectra with redshifts $\le$~0.024 since the
[O\,{\sc ii}] $\lambda$3727 \AA\ emission line falls outside of the
SDSS wavelength range.
We detect XMPGs with redshifts ranging from
0.0005 to 0.0443 and $M_g$ luminosities from $-$12\fm4 to $-$18\fm6.
Our eight new XMPGs increase the number of known metal-deficient galaxies
by approximately
one quarter.  The estimated surface density of XMPGs is 0.004 deg$^{-2}$
for $r$ $\le$ 17\fm77.
\end{abstract}


\section{Introduction}

Star-forming, extremely metal-poor galaxies (XMPGs) in the nearby Universe
have received much attention, since some of these objects may be
primeval galaxies that are now experiencing their first major
burst of star formation \citep[e.g.,][]{SS72,IT99,KO00}.
XMPGs are characterized by low
nebular oxygen abundances (12+log(O/H)~$<$~7.65; e.g., Kunth \& \"Ostlin
2000, hereafter K\"O00).  Although there have been a number of special
emission-line surveys for XMPGs
\citep[e.g., ][]{Terlevich91,Pustilnik99,Salzer2000,Ugryumov03},
very few objects of this class are known.  Combining 
measurements based on different methods from a variety of sources compiled by 
K\"O00 with the recent studies by \citet{Zee00,UM1,Pustilnik02,Skill03}
and \citet{Lee03}
gives a total of $\sim 30$ currently known or suspected XMPGs.
The majority of these galaxies
are starbursting blue compact galaxies (BCGs), while the remainder are more
quiescently evolving low-surface-brightness galaxies (LSBGs) including dwarf
irregulars \citep[see, e.g., ][]{Grebel99,Grebel00}.  Little is known
about the census, formation conditions, and evolutionary status of the
different types of XMPGs. 

We have therefore begun a systematic
search for XMPGs using the spectroscopic database
of the Sloan Digital Sky Survey (SDSS).  The SDSS \citep{York2000} is an 
imaging and spectroscopic survey that will eventually cover about one quarter 
of the sky (more than 4000 deg$^2$ have been imaged to date).
After drift-scan imaging in five bandpasses
\citep{SDSS_phot,Gunn98,Hogg01} photometry pipelines
measure photometric and astrometric properties
\citep{Lupton01, EDR02, SDSS_phot1, Pier03} and
identify candidate galaxies and quasars
for multi-fibre spectroscopy.
Each plate has  640 fibers, yielding 608 spectra of galaxies,
quasars and stars and 32 sky spectra per pointing \citep{Bla03}.
The fibers have a diameter of $3''$.  The spectra cover a
wavelength range of 3800 \AA\ to 9200 \AA\ with a resolution of $\sim1800$.
The SDSS takes spectra of a magnitude-limited sample of galaxies
\citep[mainly field galaxies brighter than $r = 17\fm77$;][]{Strauss02}
and a color-selected sample of quasars \citep{QSO02}. These include
a substantial number of emission-line galaxies.
The spectra are automatically reduced and
wavelength- and flux-calibrated
\citep[see][ and http://www.sdss.org/dr1/algorithms for more details]{EDR02,DR1}.

Owing to its homogeneity, area coverage, and depth, the SDSS provides an
excellent means of creating a flux-limited sample of XMPGs.
In this Letter we report the discovery 
of eight new XMPGs and the recovery of four previously known or suspected
XMPGs.  In Section 2 we briefly describe our analysis methods and the
determination of oxygen abundances.  In Section 3 we present our results
and discuss them.  In Section 3, our findings are summarized.
Throughout the paper a Hubble constant of H$_0$ = 75 km$\,$s$^{-1}$
Mpc$^{-1}$ is adopted.

\section{Selection of candidates and calculation of oxygen abundances}

For our work we used reduced spectra from $\sim 680$ separate plates resulting
in a total sample of $\sim 250,000$ objects with galaxy spectra.
Approximately 10\% of these plates were observed more than once, thus the
above spectra also include galaxies with multiple observations.
The total area on the sky covered by the analyzed SDSS
spectroscopic database is $\sim$3000 deg$^2$.
For the analysis of the SDSS spectra we used our own software for
emission-line data, which was
created for the Hamburg/SAO Survey for Emission-Line Galaxies
\citep[HSS-ELG;][ and references therein]{Ugryumov01}
and for the Hamburg/SAO Survey for Low Metallicity Galaxies
\citep[HSS-LM;][]{Ugryumov03}.
All emission lines were re-measured using a method detailed 
in \citet{Kniazev2000} and \citet{Pustilnik02}. The continuum was
determined with the help of an algorithm described in detail by
\citet{Sh_Kn_Li_96}. 

We extracted a galaxy sample comprised only of objects with strong
emission lines (rest frame H$\beta$ equivalent widths
EW(H$\beta) \ge$~20 \AA).
Our selection procedure for galaxies with strong emission lines
will be described in detail in \citet{Kniazev03}.
Our EW(H$\beta)$ selection
is motivated by the requirement to measure (1) the weak
[O\,{\sc iii}] $\lambda$4363 \AA\ emission line (whose intensity is usually
$I_{4363} \le 0.1  I_{H\beta}$) for a direct calculation
of the electron temperature, $T_e$, and (2)
the [O\,{\sc ii}] $\lambda$7319,7330 \AA\ lines
(usually $I_{7319+7330} \le 0.05 I_{H\beta}$) needed for the
O$^+$/H$^+$ determination. These direct measurements tend to be
possible only for spectra with sufficiently large Balmer emission
equivalent widths.

While spectra obtained with fibre-fed spectrographs may be difficult
to correct for atmospheric dispersion effects \citep{Fil82}, these 
problems are reduced for compact objects such as the 
H\,{\sc ii} regions in our target galaxies.  We find this confirmed by the
excellent agreement between our measured parameters and previously 
published values for already known XMPGs that we recovered from the SDSS 
data (see below).  The SDSS spectra presented here were taken at airmasses
of $1.18\pm0.12$, which further reduces dispersion effects.

The abundances of the ionized species and the total oxygen abundance
have been obtained with the method described by  
\citet{Izotov94,TIL95}, and \citet{IT99}, which was also used for the
galaxies from the HSS-ELG and HSS-LM projects.
The uncertainties of the measurements of line intensities, continua,
extinction coefficients, and Balmer absorption equivalent widths
have been propagated to derive the uncertainties of the element abundances.
For objects with $z \le$~0.024, the SDSS spectra do not include the
[O\,{\sc ii}] $\lambda$3727 \AA\ line,  and hence the so-called ``direct 
method'' mentioned above cannot be used.  Nor is it possible to use empirical 
R23 methods
\cite[][ etc.]{Pagel79,McCall85,Pil00}, since all of them require the 
[O\,{\sc ii}] $\lambda$3727 \AA\ line.
However, as shown and tabulated by \cite{Aller84} for $p^3$ configurations,
the measured intensities of the auroral lines 
[O\,{\sc ii}] $\lambda$7319,7330 \AA\
may be used instead to determine O$^+$/H$^+$.
The resulting O$^+$/H$^+$ values should be the same with both
methods, but since the total intensities of the
[O\,{\sc ii}] $\lambda$7319,7330 \AA\
lines are much fainter than the [O\,{\sc ii}] $\lambda$3727 \AA\ line,
the application of the auroral line method is restricted to SDSS spectra  
with sufficiently high signal-to-noise ratio.

To test the accuracy of our abundance determinations
we calculated oxygen abundances using both the 
[O\,{\sc ii}] $\lambda$7319,7330 \AA\ line method and the direct method
for spectra with $z >$~0.024.
In addition, we compared our results with those for galaxies in common with
the studies quoted in the preceding paragraph.
The published oxygen abundances of the famous XMPG I~Zw~18 
range from 7.05 to 7.2 in the northwest (NW) component along
its major axis \citep[see the Figure 5 from][]{Izot99}, and from
7.2 to 7.3 in the southeast (SE) component.
Both of these components were targeted by SDSS spectroscopy, and
our measured oxygen abundances agree within the cited uncertainties
with the previous results (see Table~\ref{tbl:main_par}).
The oxygen abundance for HS~0822$+$3542
is 7.44$\pm$0.06 according to \citet{Pust03} and is in excellent agreement
with our value of 7.45$\pm$0.02.
For HS~0837$+$4717, 12+log(O/H) = 7.64$\pm$0.04 \citep{Kniazev2000},
which agrees with our SDSS abundance of 7.62$\pm$0.02
calculated with the
[O\,{\sc ii}] $\lambda$3727 \AA\ line, and with our abundance 
of 7.63$\pm$0.03 determined using the
[O\,{\sc ii}] $\lambda$7319,7330 \AA\ lines.
Finally, for our newly discovered XMPG SDSS J0519$+$0007 we find 
12+log(O/H) = 7.46 with either method, although the value obtained using
[O\,{\sc ii}] $\lambda$7319,7330 \AA\ has a larger uncertainty.
We conclude that SDSS spectra permit accurate oxygen abundance determinations
over a studied range of 7.10 $<$ 12+log(O/H) $<$ 7.65,
and that the auroral line method
appears to yield reliable results.

\section{Results and Discussion}

\begin{figure}
\begin{center}
\epsscale{1.0}
\includegraphics[width=18cm,bb=1 340 597 765,clip=]{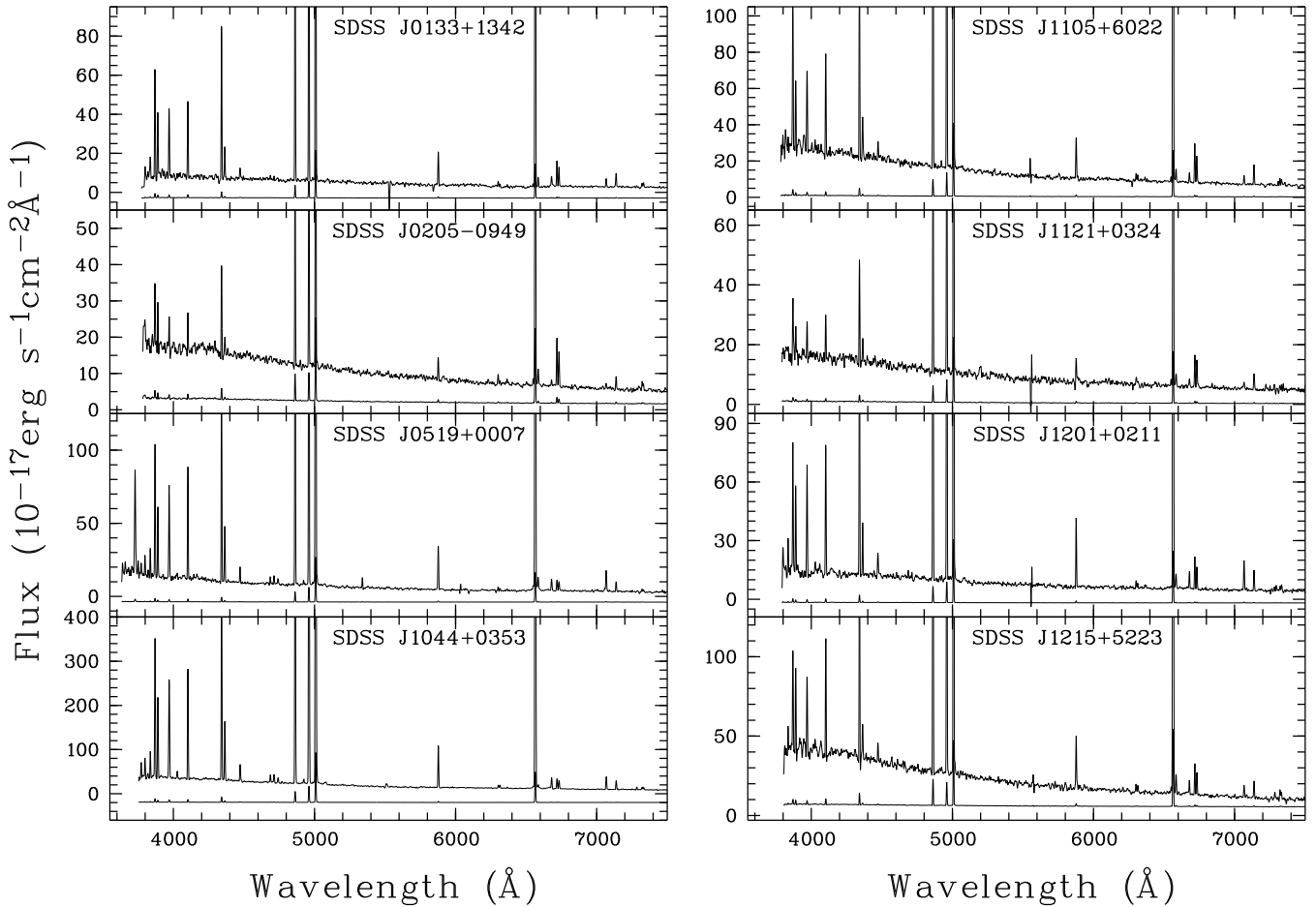}
   \caption{SDSS spectra of 8 new extremely metal-poor galaxies
	     over the rest-frame wavelength range of 3600 \AA\ to 7500 \AA.
	    The plot of the same spectrum in the bottom of each subpanel
	    is scaled to visualize the ratios of the strongest emission lines.
	   }
      \label{fig:SDSS_spectra}
\end{center}
\end{figure}

Our survey led to the discovery of eight previously unknown XMPGs.
In Figure~\ref{fig:SDSS_spectra} we show their SDSS spectra.
Table~\ref{tbl:main_par} lists the properties of the new XMPGs as well as
previously known XMPGs in the same area
that were recovered by our programs and that have an abundance
accuracy $\le 0.1$ dex.
Column 1 gives the SDSS names derived 
from the J2000 coordinates of the fiber positions.  Columns 2 and 3
contain the SDSS Petrosian integrated $g$ and $r$ magnitudes.
Column 4 lists a preliminary morphological classification following K\"O00,
column 5 SDSS redshifts $z$,
column 6 oxygen abundances 12+log(O/H) with their uncertainties,
and column 7 gives alternative galaxy
names where available from the NASA/IPAC Extragalactic Database (NED).
In Table~\ref{tbl:spectra} we present the corrected relative fluxes
of lines used for the calculation of oxygen abundances,
and their uncertainties.

In spite of three decades of searches for XMPGs, very
few such objects are currently known.  The new galaxies presented here 
increase the number of reliably known XMPGs by approximately
one quarter, an important step in efforts to establish a comprehensive
list of very metal-deficient galaxies
with well-defined, uniform selection criteria
and to study their group properties.

\begin{figure}
\begin{center}
\plotone{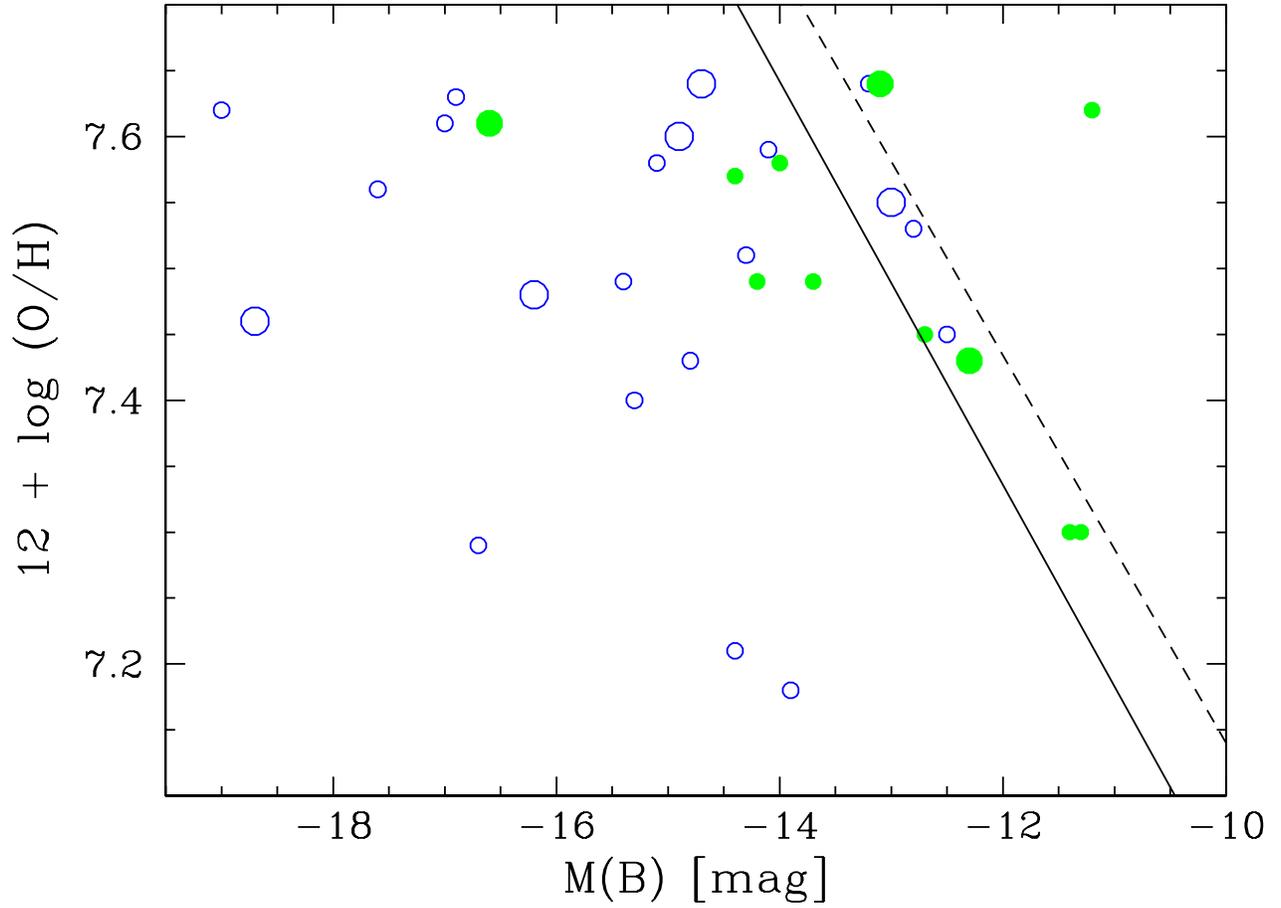}
   \caption{Blue luminosity versus oxygen abundance for previously known
            XMPGs with accurately measured oxygen abundances based on
            the direct method (small symbols; for references see Section 1), 
            and for our new XMPGs (large symbols).  Filled circles indicate
            LSBGs (mainly dIrrs).  Open circles stand for BCGs.  The
            diagonal lines indicate the luminosity-metallicity relation 
            for dIrrs (solid line: Skillman et al. 1989; dashed line:
            Richer \& McCall 1995).
	    $B$-magnitudes were recalculated from SDSS $g$ and $r$
	    Petrosian magnitudes using equations from \citet{SDSS_phot1}.
           }
      \label{fig:L-Zrel}
\end{center}
\end{figure}

Our sample of new XMPGs contains  
the most distant and most luminous
XMPG found so far (SDSS J0519$+$0007; D$\sim$177\,Mpc, M$_g\sim -18\fm6$).
SDSS J1215+5223, on the other hand, is one of the closest XMPGs if one
simply converts its redshift into distance (2.1 Mpc), although the caveat of
peculiar velocities should be kept in mind for objects with distances
of just a few Mpc \citep{Kar02,Kar03}.  The identification of very nearby 
XMPGs is of special importance since these objects permit detailed studies 
of their resolved star formation histories and of their overall age.
The XMPGs found in the SDSS data cover a wide range of abundances
(7.15 to 7.65), redshifts ($0.0005 < z < 0.0443$), and luminosities
($-12\fm4 < {\rm M}_g <  -18\fm6$).  
None of our new XMPGs is as metal-poor as I~Zw~18 (which is among the XMPGs
recovered in the SDSS data) or SBS 0335$-$052, which remain the most
metal-deficient, star-forming galaxies known to-date.
XMPGs span at least 6 magnitudes in blue luminosity, suggesting possible
atypical evolutionary histories for the most luminous objects,
since they do not follow the ``standard'' luminosity-metallicity (L-Z)
relation \citep[see, e.g., ][]{KO00}.
This is illustrated in Figure~\ref{fig:L-Zrel},
where the new and the previously known XMPGs are plotted.
While the LSBGs (primarily dwarf irregulars, dIrrs) tend to
scatter around the L-Z relation previously measured 
for dIrrs \citep{Skillman89,Richer1995}, most of the new and previously
known very metal-deficient BCGs appear to be
too luminous for their low present-day metallicities.  Interestingly,
the gas-deficient, quiescent dwarf spheroidal (dSph) galaxies would
show the opposite trend if added to this plot:  They appear subluminous
for their low metallicity as compared to dIrrs \citep[][ and references
therein]{GGH03}.  These differences are not merely a bias introduced by
the blue luminosity's sensitivity to present-day activity; in fact, the
immediate cessation of all star formation activity and subsequent passive 
fading would still require more than a Hubble time to erase the differences
in luminosity (Grebel et al.\ 2003).
 
An interesting, yet controversial scenario is the possibility that 
XMPGs that clearly deviate from the L-Z relation are
truly young galaxies, unenriched because they are now undergoing their
first episodes of star formation.  The nearest XMPGs for which detailed
stellar population studies are available are all dIrrs 
within $\sim2$ Mpc around the Local Group.  They generally
comply with the L-Z relation and 
contain old stellar populations \citep[see ][ and
references therein]{Grebel00,GGH03}.  
Deep HST imaging of nearby BCGs (typically at distances $>3$ Mpc) 
shows that star formation has progressed for at least several
Gyr \citep[e.g., ][]{Dol01,IT02}, but for more distant BCGs such as the
famous I\,Zw\,18, the age question remains unsolved
\citep[compare, e.g., ][]{Ost00,Hunt03}.  However, the older populations
in I\,Zw\,18 as well as in at least half of the new XMPGs of our sample 
should be resolvable with the Advanced Camera for Surveys aboard the 
Hubble Space Telescope, which should allow us to settle the question of
delayed star formation soon.

If BCGs underwent slow astration
for most of their history, resulting in slow enrichment similar to that
observed in dIrrs, and are
now experiencing their first major starburst, then it is conceivable that
this change in their mode of star formation
was triggered by interactions with companion galaxies, or possibly
primordial gas clouds.  
Alternatively, dwarf galaxies with particularly violent, bursty star
formation episodes may suffer significant metal loss through winds from
massive stars and supernova ejecta \citep[e.g., ][]{MF99}.
These different evolutionary scenarios result in 
predictions that can be tested via, for instance, optical, radio, 
and X-ray follow-up observations.   
An increased sample of XMPGs with well-defined, uniform selection criteria
as provided by the SDSS is crucial for uncovering their
individual and group properties.

%

By confining our search to galaxies with strong Balmer
emission we ensure that we only include galaxies for which very accurate,
direct oxygen abundance measurements based on the temperature-sensitive
[O\,{\sc iii}] $\lambda$4363 \AA\ line and on the direct calculation
of the electron temperature are possible.  With these constraints,
we find a surface density of XMPGs of $\sim 4$ per 1000 deg$^2$
for $r$ $\le$ 17\fm77 and 12+log(O/H)~$<$~7.65.
Extrapolating to the full anticipated SDSS survey area, we
expect to detect at least 20 additional galaxies once the SDSS is
completed.


\acknowledgements

LH and MAS acknowledge the support of NSF grant AST--0071079.

Funding for the creation and distribution of the SDSS Archive has been
provided by the Alfred P. Sloan Foundation, the Participating Institutions,
the National Aeronautics and Space Administration, the National Science
Foundation, the U.S. Department of Energy, the Japanese Monbukagakusho,
and the Max Planck Society. The SDSS Web site is http://www.sdss.org/.

The SDSS is managed by the Astrophysical Research Consortium (ARC) for the
Participating Institutions. The Participating Institutions are The
University of Chicago, Fermilab, the Institute for Advanced Study, the
Japan Participation Group, The Johns Hopkins University, Los Alamos
National Laboratory, the Max-Planck-Institute for Astronomy (MPIA),
the Max-Planck-Institute for Astrophysics (MPA), New Mexico State
University, University of Pittsburgh, Princeton University, the United
States Naval Observatory, and the University of Washington.

This research has made use of the
NASA/IPAC Extragalactic Database (NED) which is operated by the Jet
Propulsion Laboratory, California Institute of Technology, under contract
with the National Aeronautics and Space Administration.

\begin{deluxetable}{ccclccl}
\tabletypesize{\small}
\tablecaption{Main parameters of the new extremely metal-poor galaxies 
              discovered in the SDSS, and of the previously known, 
	      recovered ones. \label{tbl:main_par}}
\tablewidth{0pt}
\tablehead{
\colhead{SDSS Name\tablenotemark{a}} & \colhead{$g$\tablenotemark{b}} &  \colhead{$r$} &
Type\tablenotemark{c} & \colhead{$z$} & 12+log(O/H) &\colhead{Other names} \\
		    &     &  &
	      &             &           &           \\
\colhead{(1)}       & \colhead{(2)} & \colhead{(3)} &
\colhead{(4)} &\colhead{(5)}& \colhead{(6)} & \colhead{(7)}
}
\startdata
SDSS J013352.56$+$134209.4&  17.81 &17.77 & BCG & 0.00867 & 7.60$\pm$0.03 &                \\
SDSS J020549.13$-$094918.1&  15.27 &15.10 & LSBG& 0.00643 & 7.61$\pm$0.09 & KUG 0203$-$100 \\
SDSS J051902.64$+$000730.0&  18.29 &19.23 & BCG & 0.04437 & 7.46$\pm$0.02 &                \\
SDSS J104457.84$+$035313.2&  17.46 &17.67 & BCG & 0.01279 & 7.48$\pm$0.01 &                \\
SDSS J110553.76$+$602228.9&  16.37 &16.32 & BCG & 0.00447 & 7.64$\pm$0.04 & SBS 1102$+$606 \\
SDSS J112152.80$+$032421.2&  17.62 &17.38 & LSBG& 0.00383 & 7.64$\pm$0.08 &                \\
SDSS J120122.32$+$021108.5&  17.51 &17.45 & BCG & 0.00329 & 7.55$\pm$0.03 & APMUKS(BJ)     \\
SDSS J121546.56$+$522313.9&  15.75 &15.87 & LSBG& 0.00052 & 7.43$\pm$0.06 & CGCG 269$-$049 \\ \hline \\[-0.2cm]
SDSS J082555.44$+$353231.9&  17.74 &17.76 & BCG & 0.00233 & 7.45$\pm$0.02 & HS~0822$+$3542 \\
SDSS J084030.00$+$470710.2&  17.51 &18.46 & BCG & 0.04203 & 7.62$\pm$0.02 & HS~0837$+$4717 \\
SDSS J093401.92$+$551427.9&  16.03 &16.06 & BCG & 0.00253 & 7.13$\pm$0.03 & I~Zw~18 (NW component) \\
SDSS J093402.40$+$551423.3&  16.03 &16.06 & BCG & 0.00259 & 7.25$\pm$0.04 & I~Zw~18 (SE component) \\
SDSS J111934.32$+$513012.2&  16.81 &16.79 & BCG & 0.00440 & 7.51$\pm$0.04 & Arp dwarf, A1116$+$517 \\
\enddata
\tablenotetext{a}{Following IAU conventions, the official SDSS
designation for an object in the SDSS is SDSS JHHMMSS.ss+DDMMSS.s,
where the names consist of the J2000 coordinates of the source.
Hereafter we are using in the text abbreviated names, SDSS JHHMM+DDMM.}
\tablenotetext{b}{SDSS Petrosian integrated magnitude}
\tablenotetext{c}{BCG stands for blue compact galaxy.  LSBG denotes a
low-surface-brightness galaxy.}
\end{deluxetable}

\begin{deluxetable}{ccrrrrrrc}
\tabletypesize{\footnotesize}
\tablecaption{Corrected Relative Emission Line Fluxes and Errors\tablenotemark{a}. \label{tbl:spectra}}
\tablewidth{0pt}
\tablehead{
\colhead{Name}    & \colhead{[O\,{\sc ii}]}& \colhead{H$\delta$} & \colhead{H$\gamma$} & \colhead{[O\,{\sc iii}]}& \colhead{[O\,{\sc iii}]} & \colhead{[O\,{\sc iii}]} & \colhead{H$\alpha$} & \colhead{[O\,{\sc iii}]} \\
		  & \colhead{3727 \AA}     &  \colhead{4101 \AA} &  \colhead{4340 \AA} &  \colhead{4363 \AA}     & \colhead{4959 \AA}       & \colhead{5007 \AA}       & \colhead{6563 \AA}  & \colhead{7319+7330 \AA}  \\
\colhead{(1)}     & \colhead{(2)}          &  \colhead{(3)}      &  \colhead{(3)}      & \colhead{(4)}           & \colhead{(5)}            & \colhead{(6)}            & \colhead{(8)}       & \colhead{(7)}
}
\startdata
SDSS J0133$+$1342 &  \nodata               & 24.2$\pm$0.76       & 48.3$\pm$0.9        & 10.1$\pm$0.7            & 131.7$\pm$1.2            & 395.2$\pm$3.6            & 274.8$\pm$2.3       & 3.0$\pm$0.6              \\  
SDSS J0205$-$0949 &  \nodata               & 24.9$\pm$3.69       & 48.8$\pm$2.8        &  7.3$\pm$1.5            &  99.4$\pm$1.9            & 294.6$\pm$4.6            & 276.9$\pm$4.7       & 5.4$\pm$1.2              \\  
SDSS J0519$+$0007 & 33.0$\pm$1.2           & 29.4$\pm$0.54       & 47.8$\pm$0.5        & 14.3$\pm$0.3            & 142.9$\pm$0.7            & 445.6$\pm$2.1            & 274.7$\pm$1.3       & 1.1$\pm$0.3              \\  
SDSS J1044$+$0353 &  \nodata               & 25.7$\pm$0.28       & 48.0$\pm$0.4        & 13.6$\pm$0.2            & 142.4$\pm$1.2            & 443.9$\pm$3.8            & 274.6$\pm$1.7       & 1.1$\pm$0.2              \\  
SDSS J1105$+$6022 &  \nodata               & 25.8$\pm$0.78       & 47.1$\pm$1.4        & 10.3$\pm$1.0            & 138.7$\pm$1.1            & 421.9$\pm$2.9            & 277.4$\pm$2.2       & 2.8$\pm$0.4              \\  
SDSS J1121$+$0324 &  \nodata               & 26.9$\pm$2.68       & 47.1$\pm$2.7        &  8.5$\pm$1.6            & 124.7$\pm$2.6            & 352.0$\pm$6.1            & 278.0$\pm$5.1       & 4.6$\pm$1.1              \\  
SDSS J1201$+$0211 &  \nodata               & 25.6$\pm$0.62       & 48.9$\pm$0.8        &  9.7$\pm$0.5            & 124.8$\pm$1.0            & 380.5$\pm$2.5            & 277.6$\pm$1.7       & 1.7$\pm$0.3              \\  
SDSS J1215$+$5223 &  \nodata               & 25.0$\pm$1.49       & 47.6$\pm$1.4        &  6.3$\pm$0.9            &  84.5$\pm$0.8            & 236.2$\pm$2.0            & 277.0$\pm$2.6       & 3.1$\pm$0.4              \\  
SDSS J0825$+$3532 &  \nodata               & 23.6$\pm$0.35       & 44.5$\pm$0.5        & 10.3$\pm$0.3            & 119.8$\pm$0.7            & 358.9$\pm$2.1            & 275.5$\pm$1.7       & 1.4$\pm$0.2              \\  
SDSS J0840$+$4707 & 45.4$\pm$0.9           & 27.9$\pm$0.43       & 46.9$\pm$0.6        & 17.0$\pm$0.4            & 190.3$\pm$1.4            & 577.3$\pm$4.2            & 275.3$\pm$2.1       & 2.2$\pm$0.1              \\  
SDSS J0934$+$5514 &  \nodata               & 24.5$\pm$0.87       & 48.6$\pm$0.7        &  7.3$\pm$0.4            &  69.4$\pm$0.5            & 214.6$\pm$1.3            & 274.6$\pm$3.5       & 0.7$\pm$0.2              \\  
SDSS J0934$+$5514 &  \nodata               & 24.4$\pm$0.52       & 48.4$\pm$0.6        &  4.7$\pm$0.4            &  55.7$\pm$0.4            & 182.4$\pm$1.2            & 276.8$\pm$1.3       & 1.4$\pm$0.2              \\  
SDSS J1119$+$5130 &  \nodata               & 23.2$\pm$2.05       & 49.0$\pm$1.5        &  7.2$\pm$0.7            &  96.9$\pm$1.0            & 281.3$\pm$2.6            & 276.6$\pm$2.5       & 2.9$\pm$0.5              \\  
\enddata
\tablenotetext{a}{We show corrected flux ratios 100$\cdot$I($\lambda$)/I(H$\beta$)}
\end{deluxetable}


\begin{thebibliography}{}


\bibitem[Abazajian et al.\,(2003)]{DR1} Abazajian et al. 2003, submitted
(astro-ph/0305492)

\bibitem[Aller\,(1984)]{Aller84} Aller, L.H. 1984,
   Physics of Thermal Gaseous Nebulae, Dordrecht: Reidel


\bibitem[Blanton et al.\,(2003)]{Bla03}
	Blanton, M.R., Lupton, R.H., Maley, F.M., Young, N., Zehavi, I., \&
	Loveday, J. 2003, \aj, 125, 2276

\bibitem[Dolphin et al.\,(2001)]{Dol01} Dolphin, A.~E.~et al.\
   2001, \mnras, 324, 249

\bibitem[Filippenko\,(1982)]{Fil82}
  Filippenko, A.V. 1982, \pasp, 94, 715

\bibitem[Fukugita et al.\,(1996)]{SDSS_phot} Fukugita, M., Ichikawa, T.,
  Gunn, J.E., Doi, M., Shimasaku, K., \& Schneider, D.P. 1996, \aj, 111, 1748

\bibitem[Grebel\,(1999)]{Grebel99} Grebel, E.K. 1999, in IAU Symp. 192,
The Stellar Content of the Local Group, ed.\ P.\ Whitelock \& R.\ Cannon
(Provo: ASP), 17

\bibitem[Grebel\,(2000)]{Grebel00} Grebel, E.K. 2000, in Star Formation from
the Small to the Large Scale, ESA SP-445, ed. F.\ Favata, A.A.\ Kaas, \& A.\ 
Wilson (Noordwijk: ESA), 87

\bibitem[Grebel, Gallagher \& Harbeck\,(2003)]{GGH03}
Grebel, E.K., Gallagher, J.S., \& Harbeck, D. 2003, \aj, 125, 1926

\bibitem[Gunn et al.\,(1998)]{Gunn98} Gunn, J.E. et al. 1998, \aj, 116, 3040

\bibitem[Hogg et al.\,(2001)]{Hogg01}
  Hogg, D.W., Finkbeiner, D.P., Schlegel, D.J., and Gunn, J.E. 2001,
   \aj, 122, 2129

\bibitem[Hunt, Thuan, \& Izotov\,(2003)]{Hunt03} Hunt, L.~K.,
  Thuan, T.~X., \& Izotov, Y.~I.\ 2003, \apj, 588, 281 

\bibitem[Izotov et al.\,(1999)]{Izot99}
  Izotov, Y.I., Chaffee, F.H., Foltz, C.B., Green, R.F.,
  Guseva, N.G., \& Thuan, T.X. 1999, \apj, 527, 757



\bibitem[Izotov \& Thuan\,(1999)]{IT99}
  Izotov, Y.I., \& Thuan, T.X. 1999, \apj, 511, 639

\bibitem[Izotov \& Thuan\,(2002)]{IT02}
  Izotov, Y.I., \& Thuan, T.X. 2002, \apj, 567, 875

\bibitem[Izotov, Thuan \& Lipovetsky\,(1994)]{Izotov94}
  Izotov, Y.I., Thuan, T.X., \& Lipovetsky, V.A. 1994, \apj, 435, 647

\bibitem[Karachentsev et al.\,(2002)]{Kar02} Karachentsev,
  I.~D.~et al.\ 2002, \aap, 389, 812 

\bibitem[Karachentsev et al.\,(2003)]{Kar03} Karachentsev,
  I.~D.~et al.\ 2003, \aap, 398, 479 




\bibitem[Kniazev et al.\,(2000)]{Kniazev2000}
   Kniazev, A.Y., Pustilnik, S.A., Ugryumov, A.V., \& Kniazeva, T.F. 2000,
   Astronomy Letters 26, 129

\bibitem[Kniazev et al.\,(2001)]{UM1}
  Kniazev, A.Y., Pustilnik, S.A., Pramsky, A.G.,
  \& Ugryumov, A.V. 2001, \aap, 371, 404


\bibitem[Kniazev et al.\,(2003)]{Kniazev03}
    Kniazev, A.Y. et al.\, 2003, in preparation.


\bibitem[Kunth \& \"Ostlin\,(2000)]{KO00}
    Kunth, D., \& \"Ostlin, G. 2000, \aapr, 10, 1

\bibitem[Lee et al.\,(2003)]{Lee03}
    Lee, H., Grebel, E.K., \& Hodge, P.W. 2003, A\&A,  401, 141 


\bibitem[Lupton et al.\,(2001)]{Lupton01}
	Lupton, R., Gunn, J.E., Ivezi\'c, \v Z., Knapp, G.R., Kent, S.,
	\& Yasuda, N. 2001, in 
        Astronomical Data Analysis Software and Systems X, ASP Conf. Ser. 238,
	eds. F. R. Harnden, Jr., F.~A.~Primini, \& H. E. Payne
	(San Francisco: ASP), 269

\bibitem[Mac Low \& Ferrara\,(1999)]{MF99}
   Mac Low, M.-M., \& Ferrara, A. 1999, ApJ, 513, 142

\bibitem[McCall, Rybski \& Shields\,(1985)]{McCall85}
McCall, M.L., Rybski, P.M., \& Shields, G.A. 1985, \apjs, 57, 1

\bibitem[{\" O}stlin\,(2000)]{Ost00} {\" O}stlin, G.\ 2000,
   \apjl, 535, L99 


\bibitem[Pagel et al.\,(1979)]{Pagel79}
Pagel, B.E.J., Edmunds, M.G., Blackwell, D.E., Chun, M.S., \& Smith, G.
1979, \mnras, 189, 95

\bibitem[Pier et al.\,(2003)]{Pier03}
Pier, J.R., Munn, J.A., Hindsley, R.B, Hennessy, G.S., Kent, S.M.,
Lupton, R.H., \& Ivezi\'c, \v Z. 2003, \aj, 125, 1559

\bibitem[Pilyugin\,(2000)]{Pil00}
    Pilyugin, L.S. 2000, \aap, 362, 325



\bibitem[Pustilnik et al.\,(1999)]{Pustilnik99}
  Pustilnik, S.A., Engels, D., Ugryumov, A.V., Lipovetsky, V.A., 
  Hagen, H.-J., Kniazev, A.Y., Izotov, Y.I., \& Richter, G. 
  1999, A\&AS, 137, 299

\bibitem[Pustilnik et al.\,(2002)]{Pustilnik02}
 Pustilnik, S.A., Kniazev, A.Y., Masegosa, J., M\'arquez, I.M., 
 Pramskij, A.G., \& Ugryumov, A.V. 2002, \aap, 389, 779


\bibitem[Pustilnik et al.\,(2003)]{Pust03}
 Pustilnik, S.A., Kniazev, A.Y., Pramsky, A.G., Ugryumov, A.V.,
 \& Masegosa, J. 2003b, \aap, submitted.

\bibitem[Richards et al.\,(2002)]{QSO02}
    Richards, G.T., Fan, X., Newberg, H.J., et al.\, 2002, 123, 2945

\bibitem[Richer \& McCall\,(1995)]{Richer1995} Richer, M.~G.~\&
   McCall, M.~L.\ 1995, \apj, 445, 642 

\bibitem[Salzer et al.\,(2000)]{Salzer2000}
   Salzer, J.J., Gronwall, C., Lipovetsky, V.A., et al.\,
   2000, \aj, 120, 80


\bibitem[Searle \& Sargent\,(1972)]{SS72}
   Searle, L., \& Sargent, W.L.W. 1972, \apj, 173, 25

\bibitem[Shergin, Kniazev \& Lipovetsky\,(1996)]{Sh_Kn_Li_96}
    Shergin, V.S., Kniazev, A.Yu., \& Lipovetsky, V.A.
    1996, Astronomische Nachrichten, 2, 95

\bibitem[Skillman, Kennicutt \& Hodge\,(1989)]{Skillman89}
   Skillman, E.D., Kennicutt, R.C., \& Hodge, P.W. 1989, \apj, 347, 875

\bibitem[Skillman et al.\,(2003)]{Skill03}
    Skillman, E.D., C\^ot\'e, S., \& Miller, B.W. 2003, AJ, 125, 593

\bibitem[Smith et al.\,(2002)]{SDSS_phot1}
    Smith, J.A. et al. 2002, \aj, 123, 2121

\bibitem[Stoughton et al.\,(2002)]{EDR02}
	Stoughton, C.~et al.\ 2002, \aj, 123, 485

\bibitem[Strauss et al.\,(2002)]{Strauss02} Strauss, M.A.
    et al. 2002, \aj, 124, 1810

\bibitem[Terlevich et al.\,(1991)]{Terlevich91}
    Terlevich, R., Melnick, J., Masegosa, J.,
    Moles, M., \& Copetti, M.V.F. 1991, \aaps, 91, 285

\bibitem[Thuan, Izotov \& Lipovetsky\,(1995)]{TIL95}
    Thuan, T.X., Izotov, Y.I., \& Lipovetsky, V.A. 1995, \apj, 445, 108


\bibitem[Ugryumov et al.\,(2001)]{Ugryumov01}
 Ugryumov, A. V., Engels, D., Kniazev, A. Y., et al.\, 2001, \aap, 374, 907

\bibitem[Ugryumov et al.\,(2003)]{Ugryumov03}
    Ugryumov, A.V., Engels, D., Pustilnik, S.A.,
 Kniazev, A.Y., Pramsky, A.G., \& Hagen, H.-J.
    2003, \aap, 397, 463

\bibitem[van Zee\,(2000)]{Zee00} van Zee, L. 2000, \apj, 543, L31


\bibitem[York et al.\,(2000)]{York2000} York, D.G.
    et al.\, 2000, \aj, 120, 1579

\end{thebibliography}
\end{document}